\documentclass[twocolumn,showpacs,10pt]{revtex4-1}%
\usepackage{amsmath,amssymb,graphicx}
\usepackage{amsmath}
\usepackage{amsfonts}
\usepackage{amssymb}
\usepackage{graphicx}%
\providecommand{\U}[1]{\protect\rule{.1in}{.1in}}
\begin{document}
\title{Color night vision correlation imaging without an infrared focal plane array}
\author{Deyang Duan}

\affiliation{School of Physics and Physical Engineering, Qufu Normal University, Qufu 273165, China\\
Shandong Provincial Key Laboratory of Laser Polarization and Information
Technology, Research Institute of Laser, Qufu Normal University, Qufu 273165, China}
\author{Yunjie Xia}
\email{yjxia@qfnu.edu.cn}
\affiliation{School of Physics and Physical Engineering, Qufu Normal University, Qufu 273165, China\\
Shandong Provincial Key Laboratory of Laser Polarization and Information
Technology, Research Institute of Laser, Qufu Normal University, Qufu 273165, China}

\begin{abstract}
Night vision is the ability to see in low-light conditions. However,
conventional night vision imaging technology is limited by the requisite high-performance
infrared focal plane array.
In this article, we propose a novel scheme of color night
vision imaging without the use of an infrared focal plane array. In the experimental device, the
two-wavelength infrared laser beam reflected by the target is modulated by a
spatial light modulator, and the output light is detected by a photomultiplier
tube. Two infrared night vision images are reconstructed by measuring the
second-order intensity correlation function between two light fields. Thus,
the processing mode of optical electric detection in conventional night vision
imaging is transformed into the processing mode of light field control.
Furthermore, two gray images with different spectra are processed to form a
color night vision image. We show that a high-quality color night vision image can be obtained by this method.

\end{abstract}
\maketitle

\section{Introduction}

Human beings rely on a variety of senses to perceive the world, of which 83\%
comes from vision. However, in the environment of extremely weak light
intensity, human vision is significantly limited. To expand the scope of our
vision, night vision imaging technology has emerged. Night vision technology
is a kind of optical imaging that can transform an invisible scene into a
visible image by using photoelectric detection and imaging equipment under low
illumination at night. Owing to this type of technology, the scope of human
observation has now been greatly expanded. To date, night vision imaging
technology has been applied in many fields, including military reconnaissance,
security monitoring, car-assisted driving and so on. Generally, night vision
imaging technology includes thermal imaging and low-light-level (LLL) night
vision imaging [1]. The infrared focal plane array (IRFPA) is the
indispensable core device of thermal imaging. IRFPAs have high technical
requirements and poor imaging quality, preventing their wide use in commercial
applications such as visible focal plane arrays (FPAs).

The IRFPA represents one of the key problems restricting the development of
night vision imaging technology, and the performance of IRFPAs is not expected
to become comparable to that of conventional charge-coupled device (CCDs) or
complementary metal oxide semiconductors (CMOSs) in the near future.
Accordingly, is there a new scheme that can omit the IRFPA? In this article,
we propose a novel color night vision imaging scheme based on the intensity
correlation (or second-order intensity correlation) of the light field [2-6].
Photoelectric conversion in conventional night vision imaging is replaced by
the intensity correlation of the light field. Thus, the processing mode of
optical electric detection in traditional night vision imaging technology is
transformed into the processing mode of light field control. In the
experimental setup, the two-wavelength infrared laser beam reflected by the
target is modulated by a spatial light modulator (SLM), and the modulated
light is detected by a photomultiplier tube (PMT). Although an infrared light
source is used in this scheme, no IRFPA is involved in the imaging setup.

Here, light field control involves making two light fields have a second-order
spatial intensity correlation. Different from the Michelson interferometer,
which describes first-order light field properties [7], the intensity
interferometer pioneered by Hanbury Brown and Twiss (HBT), which focuses on
the second-order correlation of light intensity fluctuations, is a great
method to study optical coherence theory [8-10]. A spatial HBT-type imaging
modality, called correlation imaging (or ghost imaging) [2-6,11,12], has
attracted much attention from researchers. Compared with the classical optical
imaging approach, correlation imaging has some unique advantages, e.g.,
anti-interference [13,14] and super-resolution [15,16], indicating that ghost
imaging has important applications in lidar [17,18], remote sensing [19,20],
and pattern recognition [21].

Conventional color night vision imaging is mainly used to process gray images
output by LLL night vision and infrared night vision via software, such as
image fusion and color conversion [22-25]. In this article, according to the
intensity correlation of the light field, two gray images with different
spectra are obtained. Thus, the gray image produced by LLL night vision can be
replaced by the gray image generated by short-wavelength infrared light. Then,
the night vision image with strong sense of nature is obtained by color
conversion and gray-level fusion. Compared with the conventional pseudocolor
processing mode, this scheme does not need the LLL night vision image to
participate in pseudocolor processing. The experimental results show that this
scheme can produce a color night vision image of comparable quality to that of
the classical optical image.

\section{Theory}

We depict the scheme of color night vision correlation imaging without an
IRFPA in Figure 1. We assume that two infrared lasers with different
wavelengths $E_{s}(\omega_{1})$ and $E_{s}(\omega_{2})$ illuminate a target,
and the reflected light carrying the target's information propagates the
distance of $z_{1}$ and converges on the SLM surface through a lens. Then, the
modulated light, which propagates the distance of $z_{2}$ in free space, is
collected by a PMT and can be expressed as [21,22]%

\begin{align}
& E_{d}(x,t)=\sum_{i=1}^{2}\int d\omega_{i}dq_{i}V(q_{i})E_{s}(\omega
_{i})\nonumber\\
& \times H_{i}(x_{s},q_{i};\omega_{i})T(x_{o})H_{i}^{^{\prime}}(x_{p}%
,q_{i};\omega_{i}),
\end{align}
where the subscripts $d$ and $s$ indicate the detection light field and signal
light field, respectively. The SLM produces spatial amplitude modulation of
the light field distribution represented by the random spatial mask function
$V$, which is used to obtain spatial correlations that follow Gaussian
statistics [23]. The functions $H$ and $H^{^{\prime}}$ are transfer functions
that describe the propagation from the target to the SLM and the SLM to the
PMT, respectively. $x$ and $q$ represent the transverse position and wave
vector, respectively. $T(x_{o})$ represents the reflection coefficient of the
target. The on-target deterministic intensity pattern produced by the SLM can
then be calculated via diffraction theory:
\begin{equation}
E_{c}(x,t)=\sum_{i=1}^{2}\int d\omega_{i}dq_{i}\,V(q_{i})E_{s}(\omega
_{i})H^{^{\prime}}(x_{p},q_{i};\omega_{i}).
\end{equation}
where the subscript $c$ represents the calculation light field.
\begin{figure}[ptbh]
\centering
\fbox{\includegraphics[width=1\linewidth]{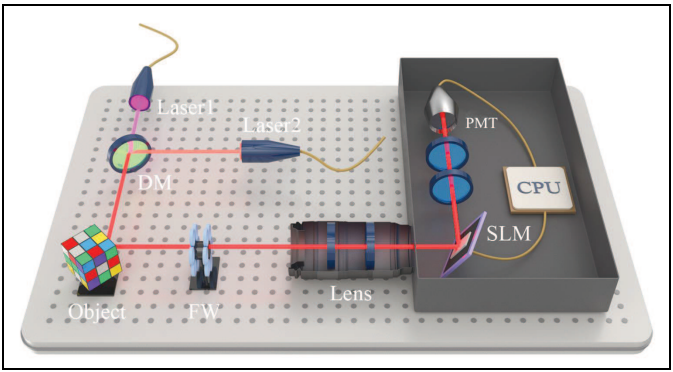}}\caption{(Color online)
Setup of color night vision correlation imaging without an IRFPA. DM: dichroic
mirror, FW: filter wheel, SLM: spatial light modulator, PMT: photomultiplier
tube. }%
\label{fig:false-color}%
\end{figure}

The image is reconstructed by the intensity cross-correlation measurement, i.e.,%

\begin{align}
&  G(x_{p},x_{s})\nonumber\\
&  =\left\langle \left\vert E_{d}(x,t)\right\vert ^{2}\left\vert
E_{c}(x,t)\right\vert ^{2}\right\rangle -\left\langle \left\vert
E_{d}(x,t)\right\vert ^{2}\right\rangle \left\langle \left\vert E_{c}%
(x,t)\right\vert ^{2}\right\rangle \nonumber\\
&  =\sum_{i=1}^{2}\int d\omega_{i}d\omega_{i}^{^{\prime}}dq_{i}dq_{i}%
^{^{\prime}}H(x_{s},q_{i}^{^{\prime}};\omega_{i}^{^{\prime}})H^{\ast}%
(x_{s},q_{i};\omega_{i})\nonumber\\
&  \times H^{^{\prime}\ast}(x_{p},q_{i};\omega_{i})H^{^{\prime}}(x_{p}%
,q_{i}^{^{\prime}};\omega_{i}^{^{\prime}})H^{^{\prime}\ast}(x_{p}%
,q;\omega)H(x_{p},q_{i}^{^{\prime}};\omega_{i}^{^{\prime}})\nonumber\\
&  \times C(\omega_{i},\omega_{i}^{^{\prime}};q_{i},q_{i}^{^{\prime}}%
{})T(x_{o})T^{\ast}(x_{o}).
\end{align}
where
\begin{align}
&  C(\omega_{i},\omega_{i}^{^{\prime}};q_{i},q_{i}^{^{\prime}}{})=\left\langle
E_{s}(\omega_{i})E_{s}(\omega_{i}^{^{\prime}})\right\rangle \left\langle
E_{i}(\omega_{i}^{^{\prime}})E_{i}(\omega_{i})\right\rangle \nonumber\\
&  \times\left\langle V(q_{i})V(q_{i}^{^{\prime}})\right\rangle \left\langle
V(q_{i}^{^{\prime}})V(q_{i})\right\rangle
\end{align}
is the intensity cross-correlation function in the spatial and temporal
frequency domain evaluated at the output surface of the SLM [23]. Substituting
Eq. (4) into Eq. (3), we can thus rewrite the ghost image as
\begin{align}
&  G(x_{o},x_{r})\nonumber\\
&  =\sum_{i=1}^{2}I_{d}I_{c}\left\vert \int dx_{r}^{^{\prime}}dx^{^{\prime}%
}W(x_{p}^{^{\prime}},x_{s}^{^{\prime}})H(x_{s},x_{s}^{^{\prime}}%
;\omega)O(x_{o})\right\vert ^{2},
\end{align}
where $I_{a}=\left\langle \left\vert \int d\omega E_{a}\left(  \omega\right)
\right\vert ^{2}\right\rangle $, with $a=d,c$ representing the product of the
average intensities of the detected light and the calculated light,
respectively. The function $W(x_{p}^{^{\prime}},x_{s}^{^{\prime}})$ is the
spatial Fourier transform of $\left\langle V(q)V(q)\right\rangle $, and the
transfer function $H$ is written in position space. We assume that
$H^{^{\prime}}=1$ because the distance ($z_{2}$) between the SLM and bucket
detector is fixed. $\left\langle T(x_{o})T^{\ast}(x_{o}^{^{\prime}%
})\right\rangle =\lambda O(x_{o})\delta\left(  x_{o}-x_{o}^{^{\prime}}\right)
$.

Through the above theoretical analysis, we obtain the following conclusions:
(1) This imaging scheme produces two infrared gray images. (2) The physical
nature of color night vision correlation imaging without an IRFPA is based on
the second-order intensity correlation of light (see Eq. (4)). (3) This
imaging scheme is not sensitive to the distance of the object and imaging
device. The distance from the SLM to the PMT is fixed. Thus, the transfer
function $H^{^{\prime}}$ is not affected by the position of the target. The
light path from the target to the SLM is adjusted by a combined lens, which is
the same as in conventional optical imaging. (4) If the SLM and PMT are
regarded as an imaging device and the combined lens is regarded as the camera
lens, this imaging scheme is exactly the same as a conventional optical camera
in terms of structure.

\section{Experiments}

Our experimental setup is illustrated in Figure 1. Two near-infrared lasers
with $\lambda_{1}=785$~nm and $\lambda_{2}=830$~nm (Changchun New Industries
Optoelectronics Technology Co., Ltd. MLL-III-785, MDL-III-830) are coupled
into a beam by a dichroic mirror (Thorlabs, DMLP805). This two-wavelength beam
in the laboratory illuminates a target, and the reflected light is converged
on the surface of a two-dimensional amplitude-only ferroelectric liquid
crystal SLM (FLC-SLM, Meadowlark Optics A512-450-850), with 512$\times$512
addressable 15~$\mu m$ $\times$15~$\mu m$ pixels. Then, the modulated light is
filtered by two bandpass filters (Thorlabs, FL05780-10, FL830-10) mounted on a
filter wheel (Daheng Xinjiyuan Technology Co., Ltd. GCM-14). Finally, the
light carrying the target's information is collected by a PMT (Hamamatsu
H10721-20). Two infrared night vision images are produced by cross-correlation
of the input signal of the SLM and the output signal of the PMT.

\begin{figure}[ptbh]
\centering
\fbox{\includegraphics[width=1\linewidth]{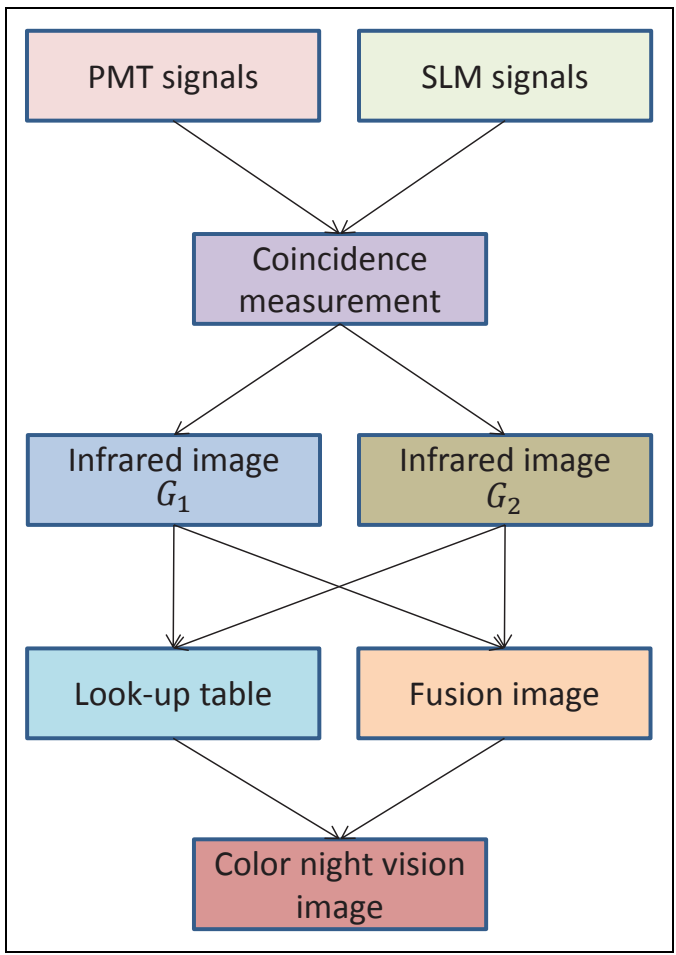}}\caption{Image processing
schematic of color night vision correlation imaging without an IRFPA. }%
\label{fig:false-color}%
\end{figure}Next, the grayscale images output by this scheme are processed
with color. Pseudocolor image processing includes two main parts: color
mapping and gray-level fusion. The flow chart is shown in Figure 2. First, a
look-up table is constructed based on a backpropagation neural network. From
the above analysis, two gray images can be obtained at the same time. These
two gray images and one color image are selected as training samples.
According to the corresponding relationship of pixel positions in different
images, the two-dimensional gray vector ($Y_{1}$, $Y_{2}$) and two-dimensional
chroma vector $(C_{b},C_{r})$ of the selected training samples are nonlinear
fitted by using a back-propagation neural network to determine the best
mapping $f(Y_{1},Y_{2})$ to $(C_{b},C_{r})$. With the help of the neural
network nonlinear fitting toolbox provided by MATLAB, the double-input and
double-output double-layer backpropagation neural network (with the hidden
layer containing 10 neurons) is constructed and trained by the
Levenberg-Marquardt algorithm. In particular, 70\% of the samples are used for
network training, 15\% are used for generalization performance optimization,
and 15\% are used for network testing. Taking $Y_{1}$ and $Y_{2}$ as
horizontal and vertical coordinates, respectively, a complete two-dimensional
color lookup table tclut is constructed by using the mapping f obtained in the
previous step and the input values ($Y_{1}$, $Y_{2}$), i.e., ($C_{b}$, $C_{r}%
$)$=T_{clut}(Y_{1},Y_{2})=f(Y_{1},Y_{2})$, where $Y_{1},Y_{2}=0,1,\ldots255$,
$(Y_{1},Y_{2})$ denotes all the two-dimensional grayscale combinations of
8-bit grayscale images, with a total of $256\times256$ $(C_{b},C_{r})$ index
values. In this way, we build a color lookup table.

For infrared gray image $G_{1}$ and infrared gray image $G_{2}$ (the pixel
size is $M\times N$ and the bit depth is 8 bits), according to the gray value
combination $(Y_{1},Y_{2})$ $(Y_{1}=G_{1}(i,j),Y_{2}=G_{2}(i,j),Y_{1},Y_{2}%
\in\lbrack0,255])$ of the two images at the same pixel position $(i,j)$
$(i\in\lbrack1,M],j\in\lbrack1,N])$, the $C_{b}C_{r}$ look-up table Tclut is
indexed. The index value tclut $(Y_{1},Y_{2})$ is the chromaticity value at
the same pixel position $(i,j)$ of color fusion image $G_{f}$, i.e.,
$[C_{b}(i,j),C_{r}(i,j)]=Gf(i,j)=T_{clut}(Y_{1},Y_{2})$. The $C_{b}$ and
$C_{r}$ chroma channel information of the color fusion image can be obtained
by indexing the pixels at all positions $(i=1,2\ldots M,j=1,2\ldots N)$ of the
two grayscale images. A color night vision image can be obtained by this
processing approach (for details, see the experimental results).

\begin{figure}[ptbh]
\centering
\fbox{\includegraphics[width=1\linewidth]{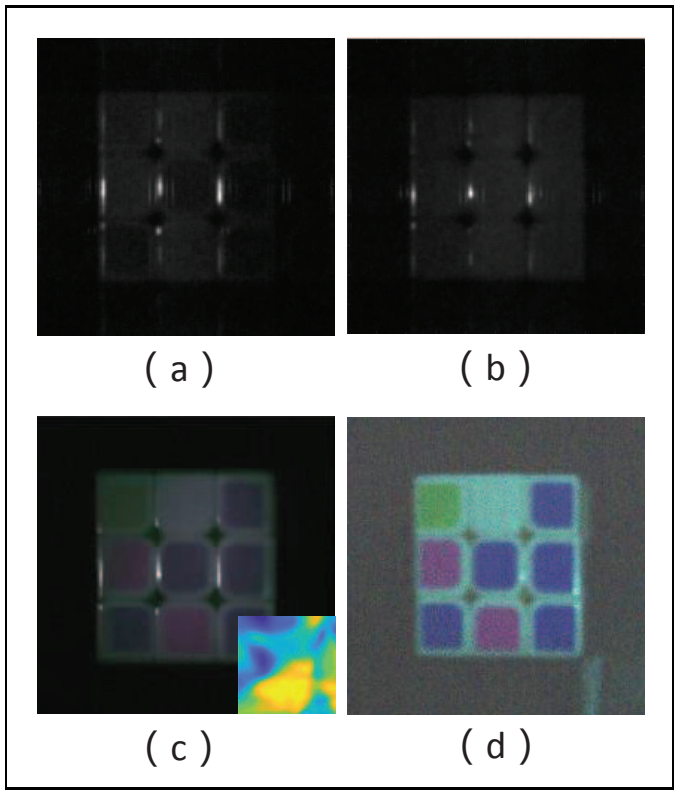}}\caption{Top row: Two
reconstructed night vision images with different wavelengths, (a)
$\lambda=785~nm$, (b) $\lambda=830~nm$. Bottom row: (c) the reconstructed
color night vision image (the illustration in the lower right corner is the
lookup table), (d) classic color image with visible light. }%
\label{fig:false-color}%
\end{figure}

\begin{figure}[ptbh]
\centering
\fbox{\includegraphics[width=1\linewidth]{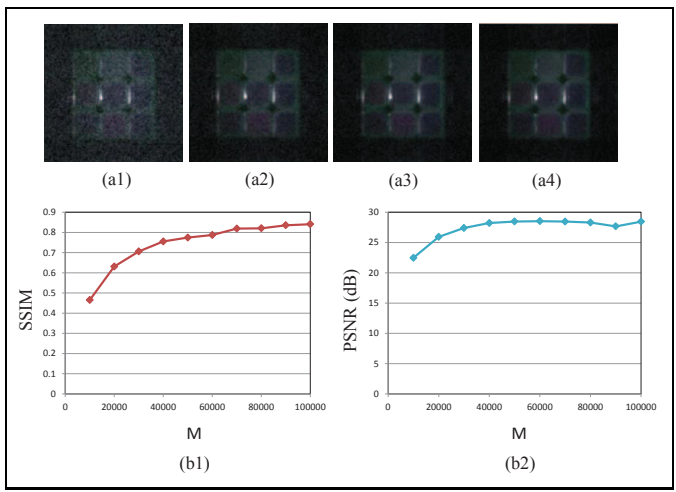}}\caption{Top row: The
reconstructed color night vision images with different realizations. The
numbers of frames are (a1) 10000, (a2) 40000, (a3) 70000, and (a4) 100000.
Bottom row: The SSIM (b1) and PSNR (b2) curves of reconstructed images with
different realizations. }%
\label{fig:false-color}%
\end{figure}In the first experiment, we chose a simple color cube as the
target. Figure 3 compares the image quality of color night vision correlation
imaging without an IRFPA, conventional near-infrared night vision and
classical color imaging with visible light. Figures 3(a,b) show the infrared
grayscale image generated by color night vision correlation imaging without an
IRFPA. Figure 3c presents the experimental results of color night vision
correlation imaging without an IRFPA with 200000 sets of data.
Correspondingly, Figure 3d depicts the classical color image with visible
light (detected by Imaging Source DFK23U618). The experimental results show
that the image obtained by our imaging scheme is of high quality and has rich
color. Compared with the near-infrared image, our image is more conducive to
scene perception and target recognition. Moreover, compared with the visible
color image, our color image does not completely restore the natural color of
the target, but the color mapping of each point is correct.

Similar to that of correlation imaging, the image quality of our imaging scheme is
related to the number of data used in image reconstruction. Figures 4(a-d)
present the experimental results with different quantities of data. To
quantitatively analyze the quality of the reconstructed image with different
numbers of data, the peak signal-to-noise ratio (PSNR) and structural
similarity index (SSIM) are used as our evaluation indexes. The results from
Figure 4 show that the image quality is significantly improved by increasing
the quantity of data.

To demonstrate that this scheme has a wide range of effectiveness, we selected
four complex color targets for the experiment. The experimental parameters are
exactly the same as those in the previous experiment. Figures 5(a1- a4) show
the four visible light images. Correspondingly, Figures 5(b1- b4) show that
the reconstructed color images exhibit high image quality and rich color.
Then, the color colorfulness index (CCI) is used to quantitatively analyze the
colorfulness of the reconstructed image. Figure 5c shows that the
reconstructed image has rich color. Although the color of the reconstructed
image cannot be completely restored to the real color of the object, the color
image is easier to distinguish and recognize than the infrared gray
image.\begin{figure}[ptbh]
\centering
\fbox{\includegraphics[width=1\linewidth]{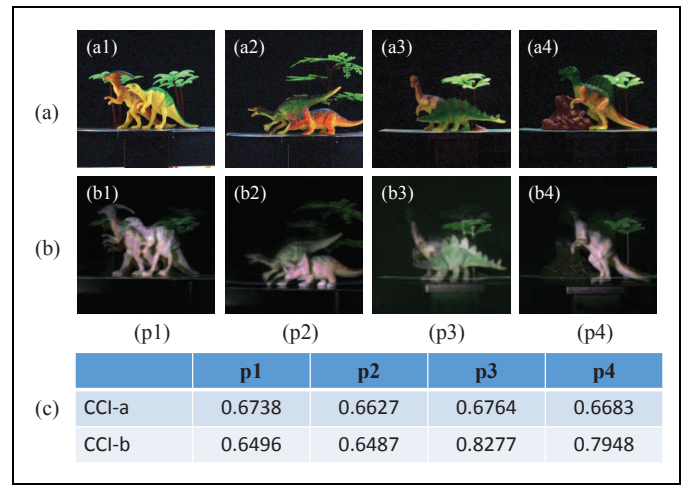}}\caption{Top row:
different targets. Middle row: The corresponding reconstructed color night
vision images. Bottom table: The CCI curves of the reconstructed image with
different targets. }%
\label{fig:false-color}%
\end{figure}

\section{Conclusion}

In this letter, we reported a novel color night vision imaging scheme without
an IRFPA. Different from conventional night vision techniques based on
first-order light field properties, our imaging scheme has a physical nature
based on second-order light field properties. The second-order intensity
correlations between the light collected by the PMT and the calculated light
are generated by modulating the SLM. The infrared night vision image can be
obtained by measuring the second-order correlation function between two
signals. The color night vision image is produced by further processing the
two infrared gray images with different spectra. Because the PMT has no
spatial resolution and the SLM does not detect the light field, the IRFPA is
not involved in the imaging process. The experimental results show that a
night vision image with rich color and high quality is obtained by color
conversion and gray-level fusion. Like other color night vision technologies,
this scheme can only partially restore the real color of the object, but the
reconstructed color image is easier to distinguish and recognize than the
traditional gray night vision image. Moreover, this scheme does not need the
LLL night vision image to participate in the pseudocolor processing. This
method provides a promising solution to improve the performance of night
vision imaging. Furthermore, this method has the potential to be developed
into a new night vision imaging technique.

\section{Acknowledgments}

This project was supported by the National Natural Science Foundation (China)
under Grant Nos. 11704221, 11574178 and 61675115, Taishan Scholar Project of
Shandong Province (China) under Grant No. tsqn201812059, and Shandong
Provincial Natural Science Foundation (China) under Grant Nos. ZR2016AP09 and ZR2016JL005.

\end{document}